\documentclass[useAMS]{mn2e}
\usepackage{lscape}
\usepackage{rotating}
\usepackage{amssymb}

\def\msun{\hbox{M$_\odot$}}
\def\mstar{\hbox{M$_\star$}}
\def\t4{\hbox{t$_{\rm 4}$}}
\def\fmid{\hbox{F$_{\rm MID}$}}

\def\cm3{\hbox{cm$^{-3}$}}

\input psfig.sty
\input epsf.sty
\usepackage{graphicx}
\usepackage{epsfig}
\title[Environmentally dependent disruption]
{Evidence for Environmentally Dependent Cluster Disruption in M83}
\author[Bastian et al.] {N. Bastian$^{1,2}$, A. Adamo$^{3}$, M. Gieles$^4$, H.J.G.L.M. Lamers$^5$, S.S. Larsen$^5$, E. Silva-Villa$^5$,  \newauthor L.J. Smith$^6$, R. Kotulla$^7$, I.S. Konstantopoulos$^8$,  G. Trancho$^9$ \&  E. Zackrisson$^3$\\
$^1$ School of Physics, University of Exeter, Stocker Road, Exeter EX4 4QL, UK\\
$^2$ Excellence Cluster Universe, Boltzmannstr. 2, 85748 Garching, Germany\\
$^3$ Department of Astronomy, Stockholm University, Oscar Klein Centre, AlbaNova, Stockholm SE-106 91, Sweden\\
$^4$ Institute of Astronomy, University of Cambridge, Madingley Road, Cambridge CB3 0HA, UK\\
$^5$ Astronomical Institute, Utrecht University, Princetonplein 5, NL-3584CC Utrecht, the Netherlands\\
$^6$ Space Telescope Science Institute and European Space Agency, 3700 San Martin Drive, Baltimore, MD 21218, USA\\
$^7$ Department of Physics, University of Wisconsin, Milwaukee, WI 53201-0431, USA\\
$^8$ Department of Astronomy and Astrophysics, The Pennsylvania State University, University Park, PA 16802, USA\\
$^9$ Gemini Observatory, Casilla 603, La Serena, Chile
}
\date{Accepted. Received; in original form}
\pagerange{\pageref{firstpage}--\pageref{lastpage}}
\pubyear{2011}
\begin{document}
\maketitle
\label{firstpage}
\begin{abstract}
Using multi-wavelength imaging from the Wide Field Camera 3 on the Hubble Space Telescope we study the stellar cluster populations of two adjacent fields in the nearby face-on spiral galaxy, M83.  The observations cover the galactic centre and reach out to $\sim6$~kpc, thereby spanning a large range of environmental conditions, ideal for testing empirical laws of cluster disruption.  The clusters are selected by visual inspection to be centrally concentrated, symmetric, and resolved on the images.   We find that a large fraction of objects detected by automated algorithms (e.g. SExtractor or Daofind) are not clusters, but rather are associations. These are likely to disperse into the field on timescales of tens of Myr due to their lower stellar densities and not due to gas expulsion (i.e. they were never gravitationally bound).  We split the sample into two discrete fields (inner and outer regions of the galaxy) and search for evidence of environmentally dependent cluster disruption.  Colour-colour diagrams of the clusters, when compared to simple stellar population models, already indicate that a much larger fraction of the clusters in the outer field are older by tens of Myr than in the inner field.  This impression is quantified by estimating each cluster's properties (age, mass, and extinction) and comparing the age/mass distributions between the two fields.  Our results are inconsistent with ``universal" age and mass distributions of clusters, and instead show that the ambient environment strongly affects the observed populations.  

\end{abstract}

\section{Introduction}
\label{sec:intro}
\vspace{-0.2cm}
The standard paradigm that has been built up over recent years is that most, if not all, stars form in clusters, and that the majority are disrupted due to the removal of the gas left over from the non-100\% star formation efficiency (e.g. Lada \& Lada 2003). This process, called `infant mortality', is thought to be largely independent of cluster mass (Hills 1980), and the timescale over which it operates is several internal crossing times of the embedded cluster. However, the importance of this infant mortality process depends crucially on how a ``star cluster" is defined (Bressert et al. 2010, Gieles \& Portegies Zwart 2011).

Beyond this early phase, theoretical and numerical studies (e.g. Spitzer~1958; Henon~1961; Baumgardt \& Makino~2003) suggest that the survival of a cluster should depend on its mass, with more massive clusters surviving longer, and environment, with strong tidal environments or high giant molecular cloud - GMC -  densities causing higher disruption rates.  However, observationally, the case is less clear, with reports of a strong dependence of disruption on cluster mass and environment (e.g. Lamers et al.~2005a,b) and no dependence on either parameter (e.g. Whitmore et al.~2007).  Quantifying the effects of disruption in cluster populations is essential in order to 1) constrain the fraction of stars that are originally formed in clusters, which in turn may affect star formation theories (e.g., Bastian~2008; Silva-Villa \& Larsen~2011), and 2) use clusters to trace the star formation histories of full galaxies or individual regions within them (e.g., Pellerin et al.~2010).

Lamers et al.~(2005a) and Gieles, Lamers, \& Portegies Zwart~(2007) have suggested that cluster populations in the Galaxy and SMC have been shaped by cluster mass and environmentally dependent disruption, respectively (see Larsen~(2006) for a thorough review of empirical mass dependent disruption laws).  In this scenario, high mass clusters are more likely to survive to older ages, and environments with low GMC densities and/or small tidal fields are more conducive to cluster survival.  Hence regions or galaxies that have a low GMC density and small tidal forces should contain a larger fraction of older clusters relative to regions with strong tidal forces.

Based on observations of the Antennae galaxies (Fall et al.~2005,2009; Whitmore et al.~2007,2010), the SMC and LMC (Chandar et al.~2006; 2010a), and the inner regions of M83 (Chandar et al.~2010b),  these authors have suggested that cluster disruption is independent of mass and environment.  They suggest that a certain fraction of clusters, \fmid, is disrupted every decade in age (with \fmid$=0.8-0.9$).  This strong cluster disruption means that the formation history of the population is a secondary effect, therefore populations should follow ``universal" age and mass distributions (although see Chandar et al.~(2011) for specific examples where this does not seem to hold within M51).  In this scenario, cluster disruption is driven largely by internal mechanisms.

Finally, Elmegreen \& Hunter~(2010) have suggested a time-dependent disruption scenario, where clusters are disrupted by the hierarchical interstellar medium from which they formed.  This is similar to the Gieles et al.~(2006) models where GMCs dominate the disruption process,  however, in the Elmegreen \& Hunter~(2010) scenario, the GMC density experienced by the cluster is time-dependent (i.e. allowing a cluster to drift away from a gas rich environment), meaning that most of the clusters are destroyed at young ages (Kruijssen et al.~2011).  This scenario predicts that cluster disruption may be independent of mass but strongly dependent on environment.


Here, we attempt to address the environmental dependence (or lack thereof) of cluster disruption, using WFC3 imaging of two adjacent   fields within M83, a nearby ($\sim4.5$~Mpc, Thim et al.~2003) face-on barred spiral galaxy.  The central pointing covers the nucleus and inner spiral arm, while the outer field has an average galactocentric distance of $\sim2$ times that of the inner field (2.5~kpc vs. 4.75~kpc, or 0.25 vs. 0.47 R$_{25}$ - Paturel et al.~2003).  Due to its smooth (although warped) outer H{\sc i} distribution and extended UV disk (Thilker et al.~2005), it is reasonable to assume that M83 has not undergone a recent strong interaction or merger, allowing us to assume a roughly constant SFR over the past $0.5-1$~Gyr, barring the inner nuclear region, which has an ongoing starburst (e.g.~Harris et al.~2001 and references therein). The assumption of a constant star-formation rate over the past $\sim1$~Gyr is the same as was made by Chandar et al.~(2010b).  Additionally, we have used the resolved stellar photometry and techniques presented in Silva-Villa \& Larsen~(2011) to derive the SFH in two fields at different galactocentric radii (centred at $\sim2.9$ and $4.8$~kpc from the galactic centre, however they are different fields than those used in the present work). We do not find an appreciable difference in the SFH over the past 100~Myr (after which the data become incomplete).  This will be discussed in more detail in a forthcoming paper.

Combining the two fields allows us to study a wide range of environmental conditions, perfect to study its effect on cluster disruption.  Chandar et al.~(2010b) have already used the inner field images to study cluster disruption in M83, and find evidence for mass and environmental independent disruption.  Here we expand on their work by considering a broader range of environments, namely by studying the inner and outer regions of the galaxy simultaneously.  Additionally, we apply the same cluster selection techniques in both fields, which allows us to carry out a comparative study, with any effects of potential selection biases minimised.  

In a forthcoming paper (Bastian et al.~in prep.) we will provide a detailed analysis of the dataset used here, along with a comparison with the catalogue of M83 clusters used by Chandar et al.~(2010b), and a more detailed discussion on cluster disruption mechanisms.  This is necessary in order to fully address the role of cluster mass in the disruption process.

\section{Observations}
\label{sec:obs}

\subsection{Data and Cluster Selection}

For the present work we use Early Release Science data of M83 (GO 11360, PI O'Connell) taken with the WFC3 on {\it HST} of two adjacent fields in M83.    The data for the inner field have been presented in Chandar et al.~(2010b), and for the current study, we only make use of the observations in the F336W (U), F438W (B),  F555W (V), F657N (H$\alpha$), F814W (I) filters.  While we use the shorthand notation for the filters, we note that no transformations were carried out.  The outer field data consist of imaging in the same filters (with the exception that the "V-band" was taken with the F547M filter), with similar exposure times, and will be presented in more detail in a future work (Bastian et al.~in prep).  The data was taken fully reduced from the WFC3 Early Release Science website\footnote{http://archive.stsci.edu/prepds/wfc3ers/m83datalist.html}.

As a first step in creating a cluster catalogue we ran {\it SExtractor} (Bertin \& Arnouts~1996) over the B, V and I images with settings chosen to select a large number of candidates.  These three catalogues were then cross-correlated, and {\it ISHAPE} (Larsen~1999) was run on the resulting list to search for resolved objects (FWHM $> 0.2$ pixels).  The V-band image of each cluster candidate was then examined, and only resolved, centrally concentrated and symmetric sources were retained as clusters.  Approximately 40-50\% of the initial candidates were removed during this step. The vast majority ($>98$\%) of the objects remaining in our sample after this step satisfy the definition of a cluster suggested by Gieles \& Portegies Zwart~(2011), namely that the age of the cluster is longer than a crossing time (based on radius and mass estimates).  While this method of cluster selection naturally introduces a certain level of subjectivity, we found that it was necessary in order to remove unbound (i.e. filamentary) groups/associations and chance alignments of two or more bright stars. See Silva-Villa \& Larsen~(2011) for a more thorough discussion of selecting clusters.  Details of the sample selection will be presented in a future work, however we note that the same selection procedure was used in both fields, so any differences should be physical and not due to any possible selection biases\footnote{The full catalogues of observed and estimated cluster properties can be found on the CDS website.}. 

We then removed all clusters within 450~pixels ($\sim400$~pc) of the galactic center, as the star formation rate is likely to have been highly variable there (e.g. Harris et al.~2001), and the detection limit is significantly worse than in the surroundings.  Aperture photometry was then carried out on all the images with an aperture, inner and outer background annuli or 5, 8, \& 10~pixels, respectively.  Photometric zeropoints were taken from the STScI website, with an additional 4\% efficiency in the F336W filter (priv. comm. Jason Kalirai) taken into account due to better than expected efficiency.  Aperture corrections for each filter were derived using a list of $\sim15$ resolved clusters common to all filters.  Comparison between our photometry and that of Whitmore et al.~(2011, hereafter W11) shows good agreement (deviations in colour $< 0.03$~mag, accounting for the F336W shift, which was not applied in W11).

\subsection{Estimating Cluster Parameters and Comparison with Previous Works}
\label{sec:sed_fitting}

We estimated the age, mass, and extinction of each of the clusters by comparing the observed cluster magnitudes to simple stellar population (SSP) models.  Two methods were used, in the first we adopted the {\it GALEV} SSP models (Kotulla et al.~2009) of 2.5 times solar metallicity\footnote{We adopt 2.5 times solar metallicity for both fields, as the metallicity gradient in M83 is such that we only expect a change in metallicity of  $\Delta(12 + {\rm log}(O/H)) \approx 0.2$  in the mean of the two fields (Kewley et al.~2010).}, and a Kroupa IMF and used the {\it 3DEF} fitting code (see Bastian et al.~2005).  In the second method, we adopted the fitting procedure of Adamo et al.~(2010a,b) and the {\it Yggdrasil} SSP models (Zackrisson et al.~2011),  adopting the same stellar population parameters as the {\it GALEV} models.  Both sets of models include nebular emission (e.g. H$\alpha$, [O{\sc iii}]) which is helpful in distinguishing old clusters from young, highly extincted clusters (e.g. Chandar et al.~2010b; Konstantopoulos et al.~2010) and contributes to the broad-band colours.  Both methods gave consistent ages/masses for the clusters. In the present work we adopt the ages/mass derived using the Adamo et al.~(2010a) method\footnote{ In a future work (Adamo et al.~ in prep) we will compare the results in more detail, and investigate systematic effects based on the combination of filters and models used (e.g. Adamo et al.~2010b; Reines et al.~2010).  }.

We limit our analysis, when cluster parameters are fit, to those clusters that have masses in excess of $5\times10^{3}$\msun\ in order to minimise stochastic sampling effects that can severely affect the derived age/mass distributions (e.g. Ma{\'{\i}}z 
Apell{\'a}niz~2009; Silva-Villa \& Larsen~2011).  After applying this limit (and avoiding the inner $\sim450$~pc of the galaxy) our final sample contains 381 and 370 clusters in the inner and outer fields, respectively.

\begin{figure}
\center
\includegraphics[width=8cm]{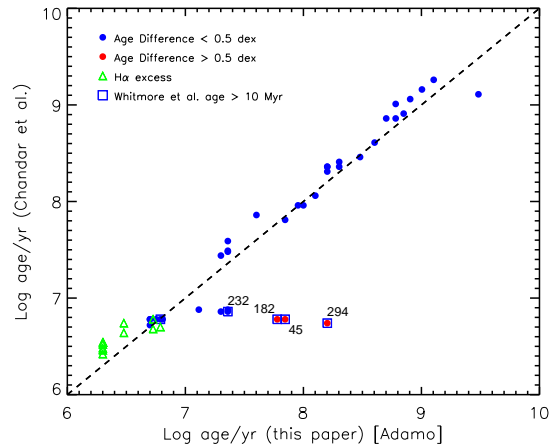}

\caption{A comparison of the estimated ages of clusters in the Chandar et al.~(2010b) sample with those presented in this paper (adopting the ages from the Adamo et al.~(2010a) method).  Note that only clusters appearing in the W11 catalogue are shown.  The dashed line shows a one-to-one correlation.  The clusters marked with squares are those for which W11 assign older ages (their category 5a or higher) based on H$\alpha$ morphology.  For clusters 45, 182, 232, and 294 our derived ages match those estimated by W11.}
\label{fig:age_comp}
\end{figure}

In Fig.~\ref{fig:age_comp} we compare the ages we derived (using the {\it Adamo} models) vs. the modelled ages presented in W11 (taken from the catalogue of Chandar et al.~2010b).  Overall, the agreement is extremely good.  The three outlying clusters, where we assign significantly older ages than derived by Chandar et al.~(2010b) all lack significant H$\alpha$ emission, and their morphology suggests older ages (W11).  

\section{Tracing Cluster Disruption}
\label{sec:disruption}

To address the role of environment in cluster disruption we have used two methods, which are presented in this section in order of their dependence on models.  The first, and least-model dependent, uses the observed colour distributions directly.  The second uses the ages and masses estimated through comparison with SSP models (see \S~\ref{sec:sed_fitting}) to study how the age and mass distributions differ as a function of environment.  

\subsection{Colour-Colour Diagrams}
\label{sec:disruption1}

\begin{figure}
\center
\includegraphics[width=8cm]{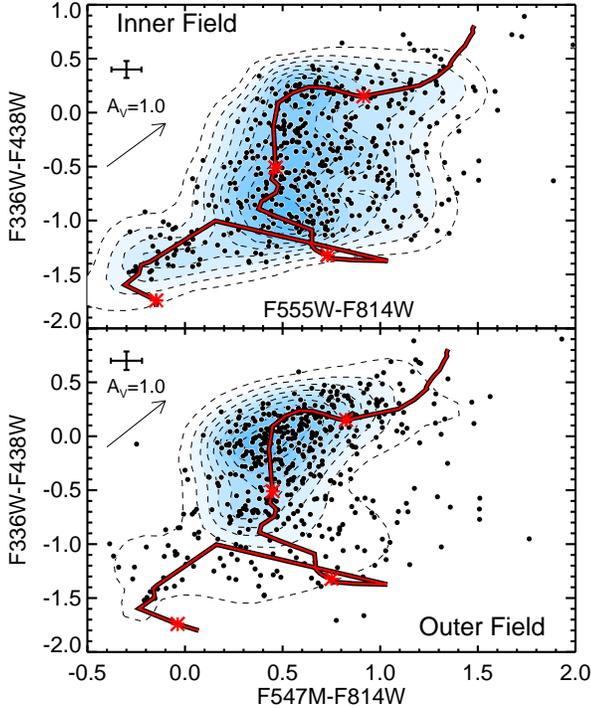}
\caption{Colour-colour diagrams for the inner (upper panel) and outer (lower panel) fields.  Each cluster is shown as a filled circle, while the contours denote the number density of points at that colour.  The (red) solid lines are the SSP model tracks used in Adamo et al.~(2010b).  For reference the models are marked with asterisks at 1, 10, 100 \& 1000~Myr, from lower-left to upper-right.  Note the differences in the distributions, with a significantly higher fraction of clusters in the outer field having colours consistent with being older.   The average error in colour for our cluster sample is shown in the upper left of each panel, and the extinction vector is also shown.}
\label{fig:colours}
\end{figure}

In Fig.~\ref{fig:colours} we show the (U--B) vs. (V--I) colours of all clusters in the inner and outer fields.  Note that the V-band corresponds to the F555W and F547M bands in the upper and lower panels, respectively.  However, the y-axis uses the same combination of filters, and hence this is where we will concentrate our analysis.  A clear difference between the two fields is the larger fraction of clusters in the outer field that have redder (U-B) colours, while the (V--I) colour distributions are comparable.  The significance of this result is easily seen when SSP models are overplotted on the data (red solid lines), which span from young ($<10$~Myr) in the lower left to old ($>$ 1~Gyr) ages in the upper right.  Hence it is already clear from this plot that the average age of the clusters is older in the outer field, suggesting that the cluster population this field is less affected by disruption.  This conclusion is independent of any assumptions on cluster metallicity.

In principle, this result could also be due to a higher extinction in the outer field, but this is not a favoured scenario as 1) the inner regions host more gas/dust per unit surface area (e.g., Crosthwaite et al.~2002) and 2) we do not see as many reddened clusters ($V-I > 0.8$) in the outer field.  In Fig.~\ref{fig:colours} we are showing all detected clusters in each field, although we note that mass and luminosity limited samples show the same distributions.  

\subsection{Age and Mass Distributions}
\label{sec:disruption2}

\begin{figure}
\center
\includegraphics[width=8cm]{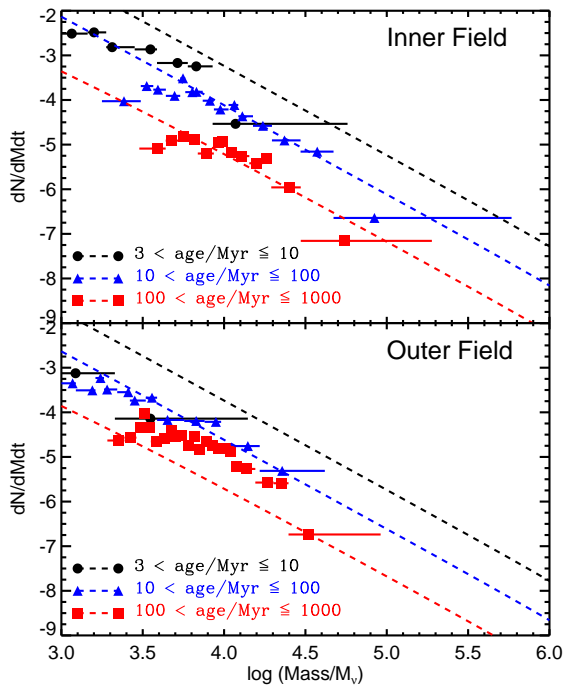}
\caption{The mass distributions for three age bins (normalised to the linear age range considered) for the inner (top panel) and outer (bottom panel) fields.  The points represent the median mass in each mass bin, while the horizontal lines show the mass range covered by the bin.  The dashed lines show the predicted ``universal" cluster distributions of the mass independent disruption scenario ($\fmid=0.9$) normalised to fit the 10-100~Myr distributions.  We have used bins with equal numbers of clusters (10 per bin).  The distributions in the upper and lower panels are clearly different, suggesting that disruption has been much less pronounced in the outer field than in the inner field.}
\label{fig:dndmdt}
\end{figure}

Using the estimated ages and masses from \S~\ref{sec:sed_fitting}, we can directly compare the age/mass distributions between the two fields.  We chose to follow the method of de Grijs \& Goodwin~(2008), also used by Chandar, Fall, \& Whitmore~(2010a) and Silva-Villa \& Larsen~(2011), namely to split each sample into three age bins, and plot the mass functions normalised to the linear age range of each bin.  The results are shown in Fig.~\ref{fig:dndmdt}.  Note that the mass functions extend to higher masses in the inner field.  This is likely due either to size-of-sample effects (e.g. Hunter et al.~2003, Gieles \& Bastian~2008) or through a higher upper cut-off in the mass function, i.e. \mstar\ if the mass function is described as a Schechter function (Larsen~2009, Gieles~2009).  

If there was no cluster disruption we would expect each of the three distributions in each panel to lie on top of each other.  However,  since the distributions are separated it shows that either the cluster-formation history has been increasing over the timespan shown (e.g. as may be expected in starburst galaxies, see Bastian et al.~~2009), or more likely in this case, disruption has been acting on the population.  If the differences between the distributions were due entirely to an increasing SFH over the past Gyr, an increase of a factor of 10 and 4 would be required in the inner, and outer fields, respectively.

Additionally, we overplot the expected ``universal" cluster distributions of Whitmore et al.~(2007), Fall et al.~(2009), and Chandar et al.~(2010a,b) as dashed lines, normalised to the $10-100$~Myr age bin.  For this, we adopted a mass-independent fraction of clusters that are destroyed each decade in time, choosing \fmid$=0.9$, and a cluster mass function of $N(dM) \propto M^{\alpha}$dM, with $\alpha=-2$.  We have chosen to normalise the ``universal" distributions to the $10 < $ age/Myr $ \le 100$, in which case these distributions over and under-predict the younger ($\le 10$~Myr) and older ($>100$~Myr) distributions, respectively.

\vspace{-0.cm}
Comparing the upper and lower panels in Fig.~\ref{fig:dndmdt} we see that the distributions of the three age bins are vertically spaced further apart in the inner field.  The fact that the observations lie nearly on-top of each other in the outer field suggests that disruption has had a relatively small effect on the cluster population far from the galactic centre.  Additionally, we have carried out the same analysis for both fields,  only using clusters with ages $<100$~Myr (i.e. where the derived SFH based on the resolved stellar population is known to have been constant, see \S~1), and come to the same conclusion. The observed differences between the two fields clearly shows the environmental dependence of cluster disruption, with disruption affecting the inner field more than the outer field.




\section{Discussion and Conclusions}
\label{sec:discussion}

Using two methods, we have shown that cluster disruption is dependent on the galactocentric distance (i.e. the ambient environment including GMCs and the tidal field)  in M83.  In the inner regions of the galaxy, where tidal effects are stronger and the density of GMCs is higher, we find evidence for higher levels of cluster disruption relative to the outer field.  This result is based on a purely empirical comparison of the cluster mass functions versus age in the two fields, and does not depend on any specific theoretical assumptions about the physical mechanism behind cluster disruption. Our results could also be explained by differing cluster formation histories between the two fields, however  the derived SFH based on the resolved stellar population does not support this scenario, at least for clusters that formed during the past 100~Myr.


The observed differences between the inner and outer fields is in agreement with the mass and environmentally dependent disruption model of Lamers et al.~(2005b), as the gas surface density in the inner region is higher by a factor of $3-4$ than the outer region (Crosthwaite et al.~2002).  The observed differences also agree, at least in a qualitative sense, with the mass-independent cluster disruption model by Elmegreen \& Hunter~(2010) as well as the mass-dependent model presented by Gieles et al.~(2006), as the ambient GMC density is expected in both models to strongly affect the disruption process.  However, the ``universal" distribution (e.g. Whitmore et al.~2007; Fall et al.~2009; Chandar et al.~2010a,b), where mass and environmentally independent cluster disruption is the dominant factor in the population,  clearly does not fit the data.  

In a future work, we will compare our catalogue of clusters in the inner field more thoroughly to that presented in Chandar et al.~(2010b) in order to quantify the effects of cluster selection.  This is necessary to fully address the role of mass in the cluster disruption process.  Additionally, we will analyse the mass distributions of clusters and the results of the full three dimensional model fitting (\t4, \fmid, \mstar) to the observed distributions.


\section*{Acknowledgments}

We thank Barbara Ercolano for helpful discussions on model fitting and Max Mutchler for his help with the images used in this work.  This paper is based on observations taken with the NASA/ESA {\it Hubble Space Telescope} obtained at the Space Telescope Science Institute, which is operated by AURA, Inc., under NASA contract NAS5-26555. The paper makes use of Early Release Science observations made by the WFC3 Science Oversight Committee.

\bsp
\label{lastpage}
\end{document}